\documentclass[a4paper,10pt]{scrartcl}

\usepackage[latin1]{inputenc}
\usepackage{lmodern}            
\usepackage[T1]{fontenc}
\usepackage[dvips]{graphicx}
\usepackage{url,paralist}
\usepackage[small,it,bf,hang]{caption2}
\usepackage{geometry}
\newcommand{\vi}{{\sffamily VILAB}}
\newcommand{\FU}{Fern\-Universität in Hagen}
\author{E. Braun, R. Lütticke, I. Glöckner, H. Helbig}
\title{Interactive Problem Solving in Prolog}
\begin{document}
\maketitle
\begin{center}
\FU\\
Praktische Informatik VII\\
Intelligente Informations- und Kommunikationssysteme (IICS)\\
Universitätsstr. 1, 58084 Hagen\\
\texttt{erik.braun@fernuni-hagen.de}
\end{center}

\begin{abstract}
\noindent \textbf{Abstract:}
This paper presents an environment for solving Prolog problems which has been implemented as a module for the virtual laboratory \vi . During the problem solving processes the learners get fast adaptive feedback. As a result analysing the learner's actions the system suggests the use of suitable auxiliary predicates which will also be checked for proper implementation. The focus of the environment has been set on robustness and the integration in \vi .
\end{abstract}
\section*{Introduction}
The Intelligent Information and Communication Systems group at the distance university \FU\ has developed a virtual laboratory (\vi\footnote{Homepage: \url{http://pi7.fernuni-hagen.de/vilab/}

\noindent Guided Tour: \url{http://inflabor.fernuni-hagen.de/tour_en/}}) for computer science focused on problem solving \cite{Lütt04}. It supports interactive problem solving in computer science with access to complex software-tools. The problems given in \vi\ are divided into different domains of
computer science (e.g. relational databases, programming, natural language processing). The learner's solutions for these problems can automatically be analysed and a tutoring component immediately gives feedback for improvements. After this the learners modify their solution and a further analysis starts, etc. \vi\ provides a learner model, which consists among other things of domain specific knowledge of the student, a distance between the students solution and the correct solution and the change of this distance for each task.

This paper presents a mechanism for analysing solutions in the field of Prolog programming. In the first part of the paper we explain how the solutions are judged, in the second part we show how the analysing component is implemented, and finally we give an outlook on further work.

\section*{Analysis of Prolog Programs}

The analysis of a solution is done in several steps:
\begin{compactenum}[(1)]
\item At first, a syntax check is performed. Compared to the regular Prolog compiler, it produces error messages which are easier to understand for Prolog beginners. In addition, it also explains how to correct typical syntax errors.
\item In the next step, the program is checked for so-called »undesirable commands«. This is necessary, since the full set of Prolog commands is made available to the learners, but some commands for the interaction with the operating system or the file system can be used (unintentional) in a destructive way. Due to Prolog's feature of dynamic command creation, the check for undesirable commands takes place also during the program run.
\item The semantic checks examine the behaviour of the learner's solution. They require a sample solution which behaves as requested. A flexible and easily extensible set of test cases shall ensure,
that all eventualities are checked. The dynamic generation of test cases makes it difficult to outsmart the analysing component by providing test case solutions as facts. In case of a very large or infinite solution set, just an inital part is generated and examined.
\item Optionally a structure check can take effect. This test can scan simple programs in detail
and compare them to a target solution, e.~g.~Euclid's GCD algorithm. On the other hand the structure check can test single predicates of a suggested solution process in more extensive tasks.
\end{compactenum}
If one of the first two checks fails, the analysis of the learner's solution aborts. In the other two cases the learner will be informed about the unexpected behaviour and asked for a correction. If all checks succeed, the solution is marked as correct.

The learner will be informed about a suggested procedure of solution by degrees, depending on the number of unsuccessful attempts.

\section*{An Example Session}

The students start their Prolog problem solving session like every other
task in \vi\ via the so-called \textit{Labortool}. They choose an exercise
and get a comprehensive description of the problem to be solved. The
learners enter the solution into a text area and submit it then to the \vi\ server. For shortness, let us assume the problem is:
 
 \smallskip
 \hspace*{1cm}\textsl{Write a program, that finds the next prime larger than a given number.}
 \smallskip
 
After an adjustable number of wrong submissions, the correction script can give several helping hints, e.~g.\ suggestions for algorithms or code fragments. If a suggested algorithm is assembled from more than one predicate, every predicate can be checked of its own. In our example, the following hint is given:
 
 \smallskip
\hspace*{1cm}\textsl{Your program may be constructed in this way: for finding the \texttt{next\_prime} you can check numbers individually with a predicate \texttt{is\_prime/1} if it is prime. If you write this predicate, it will be checked. Implementing \texttt{is\_prime/1} becomes easier if you use a predicate \texttt{has\_factors/1} which checks if a number has factors. If you write this predicate, it will be checked.}

\smallskip

Now, if the student submits an incorrect program, the correcting script checks the suggested predicates and by this way isolates the wrong parts of the program.

\smallskip
The arithmetical example was chosen to show a concise task. Problems have been created as well as correcting scripts for the practice in programming with data structures and pattern directed programming.

\section*{Implementation}
\begin{figure}[ht]
\centering
\includegraphics[width=0.9\textwidth,height=11cm]{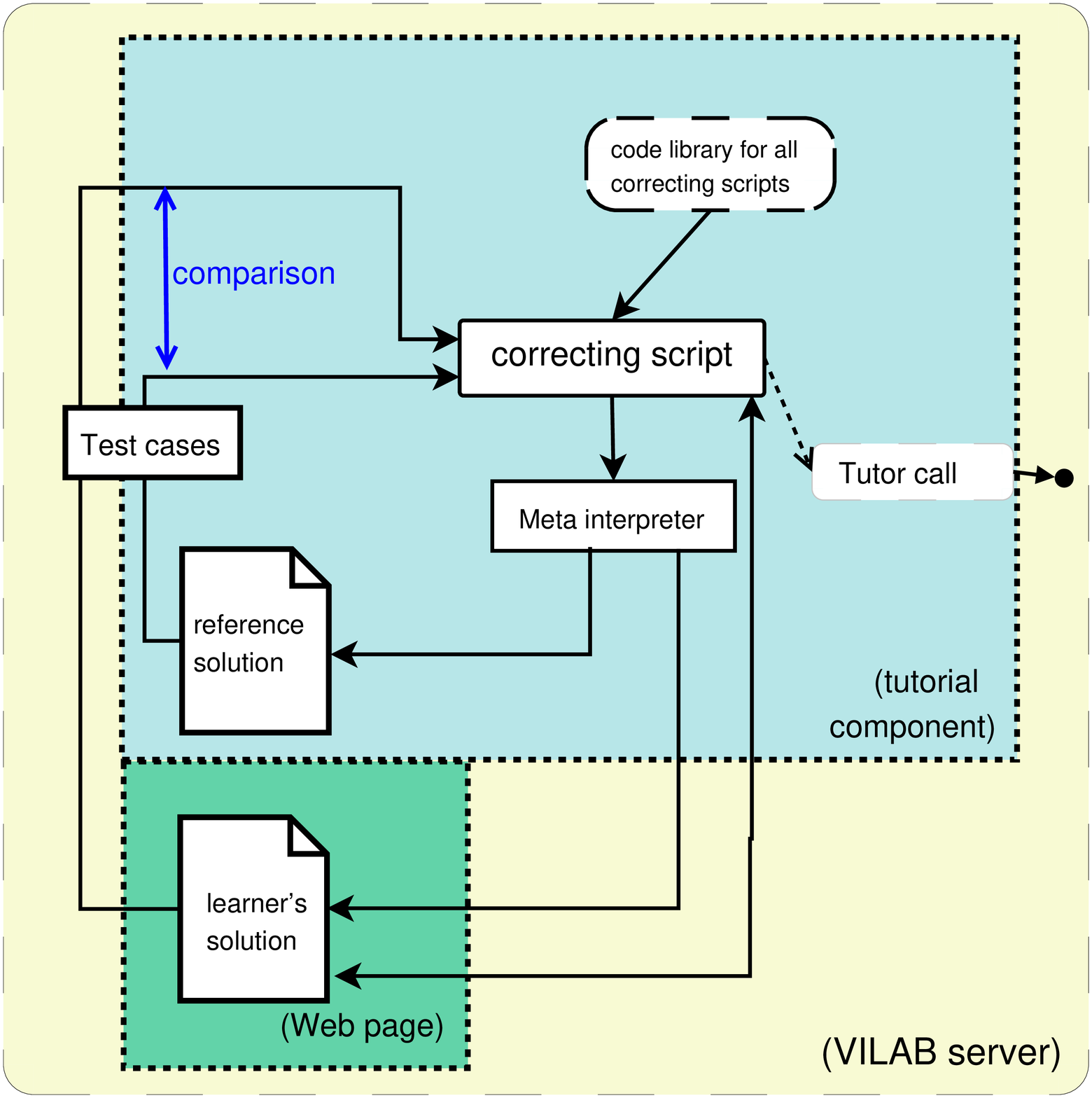}
\caption{Program Structure}
\label{fig:structure}
\end{figure}

The tutorial component is completely integrated in \vi . As shown in Figure \ref{fig:structure}, each problem consists of a Webpage, which describes the learner's task, and a corresponding correcting script, that makes extensive use of a program library providing the necessary functions for comparing data and running Prolog programs. In addition, every task needs its own reference solution and the set of test
cases. The central part is the correcting script written in Perl, which is called by the learner when sending in the solution. The correcting script starts the student's and the reference Prolog programs via a meta-interpreter to keep track of the Prolog command calls and collects the results. Rather than using static test cases, the system introduces a syntax for dynamically specifying test cases which will be randomly chosen in the specified argument range. The results are evaluated and sent back to the web page, possibly with additional hints. At the same time, the learner model in \vi\ is updated.

Due to the separated running environment, an incorrect submission can not affect the correcting script.

\section*{Further work}

There exist some packages for analysing Prolog programs with different approaches. \cite{Hong98} and \cite{Gust99} focus on teaching Prolog, whereby the learner ist restricted in commands und programming techniques he can use. Other implementations as \cite{Beie05} provide the full Prolog language, in this work the focus is targeted on program correctness by fully automated pre-testing of Prolog programming assignments.

This present work evolved from creating Prolog problems for students, and consequently in the future additional problems will be provided.
Specifically, exercises related to a Prolog-based NLP course taught at the distance university will be added.

Due to the integration in \vi\ the learners use an uniform environment they are, in contrast with insulated Prolog tutors, already used to.

The usage of all features of the learner model in \vi\ has not been finished yet; an algorithm for computing the distance between the learners submission and a correct solution should be developed.

\end{document}